\def\year{2018}\relax
\documentclass[letterpaper]{article} 
\usepackage{aaai18}  
\usepackage{times}  
\usepackage{helvet}  
\usepackage{courier}  
\usepackage{url}  
\usepackage{graphicx}  
\usepackage{pgfplots}
\usepackage{filecontents,pgfplots}
\usepgfplotslibrary{dateplot}

\begin{document}
%
\title{For Whom the Bell Trolls: Troll Behaviour in the Twitter Brexit Debate}
\author{
Clare Llewellyn, Laura Cram, Adrian Favero, Robin L. Hill\\
School of Social and Political Science\\
Crystal Macmillan Building, 15a George Square\\
Edinburgh, UK, EH8 9LD\\
{C.A.Llewellyn@sms.ed.ac.uk, Laura.Cram@ed.ac.uk, A.Favero@ed.ac.uk, r.l.hill@ed.ac.uk}
}
\maketitle
\begin{abstract}
In a review into automated and malicious activity Twitter released a list of accounts that they believed were connected to state sponsored manipulation of the 2016 American Election. This list details 2,752 accounts Twitter believed to be controlled by Russian operatives. In the absence of a similar list of operatives active within the debate on the 2016 UK referendum on membership of the European Union (Brexit) we investigated the behaviour of the same American Election focused accounts in the production of content related to the UK-EU referendum. We found that within our dataset we had Brexit-related content from 419 of these accounts, leading to 3,485 identified tweets gathered between the 29th August 2015 and 3rd October 2017. The behaviour of the accounts altered radically on the day of the referendum, shifting from generalised disruptive tweeting to retweeting each other in order to amplify content produced by other troll accounts. We also demonstrate that, while these accounts are, in general, designed to resemble American citizens, accounts created in 2016 often contained German locations and terms in the user profiles.
\end{abstract}

\section{Introduction}

Trolls are internet users who attempt to manipulate opinion by spreading rumours, speculation and false information \cite{mihaylov2015finding}. Twitter identified 2,752 troll accounts they claim are likely run by the Internet Research Agency (IRA), a Russian company that was identified as tweeting about the US 2016 elections.\footnote{\url{https://democrats-intelligence.house.gov/uploadedfiles/exhibit_b.pdf}} \footnote{Twitter released a further 1,062 accounts on the 19th January 2018, these are not included in this study}

We have been collecting tweets on the topic of the UK-EU `Brexit' referendum since August 2015. In our dataset we have 3,485 tweets from 419 troll accounts that were collected between the 29th August 2015 and 3rd October 2017. These tweets were about the Brexit vote and topics that were expected to influence the vote, such as the EU, refugees and migrants. In total, we have collected over 70 million Brexit-related tweets; 3,485 is therefore a tiny proportion of the overall number but does indicate that these trolls, who tweeted about the US elections, were also active in the Brexit debate.

As a consequence of our ethical procedure and following legal advice we are unable to disclose or share usernames, the usernames of retweeted users (unless they are verified users) or any full tweets. We utilise this information in our analysis but it remains confidential. Access to any images or videos contained in these tweets is no longer possible as these have been removed from the Twitter website. Our findings, utilising the tweet text and metadata from derived and aggregated data allow us to provide important insights into the behaviour of these Twitter trolls and the types of content that their tweets contain.

\section{Senate Testimony}
On October 31st 2017 Sean Edgett, a legal representative of Twitter, presented evidence to the United States Senate Judiciary Subcommittee on Crime and Terrorism. \footnote{\url{https://www.judiciary.senate.gov/download/10-31-17-edgett-testimony}} His report provided details of  36,746 accounts that automatically generated election content (referred to as `bots'). He also provided details of 2,752 accounts that were linked to the IRA, a Russian company thought to be involved in the creation of online propaganda. He stated that these 2,752 accounts were identified using information obtained by Twitter from third-party sources. These 2,752 accounts also produced automated content, but approximately 53\% of the content associated with these accounts was produced by humans (referred to as `trolls').  

Twitter studied tweets from 1st September 2016 to 15th November 2016. Not all of the content posted by these accounts during the time period studied was about the US election (only approximately 9\%) and over 47\% of the content was automated. In written testimony these accounts were described as being `Russian election-focused efforts'.\textsuperscript{2} The troll accounts posed as news outlets, activists, and politically engaged Americans. Edgett's testimony describes the troll behaviour as: contacting prominent individuals through mentions, organising political events and abusive behaviour and harassment. All 2,752 accounts have been suspended and the information posted by them is therefore no longer available through Twitter.

As part of a review of `Fake News' the British Member of Parliament, Damian Collins, who is the Chair of the Digital, Culture, Media and Sports Select Committee asked that the UK parliament be provided with `a list of accounts linked to the Internet Research Agency and any other Russian linked accounts that it [Twitter] has removed and examples of any posts from these accounts that are linked to the United Kingdom'. \footnote{\url{http://www.parliament.uk/documents/commons-committees/culture-media-and-sport/171103-Chair-to-Jack-Dorsey-Twitter.pdf}} Twitter responded with six tweets from Russia Today. In the absence of a specific officially-published list, detailing accounts from the IRA that were active in the Brexit debate, we investigate whether any of the accounts known to be active on the 2016 US Election also produced content related to Brexit. 

\subsection{Tweets and Retweets}

In the report given by Twitter to the Senate committee\textsuperscript{2} it is stated that, of the tweets studied from the 1st September 2016 to the 15th November 2016, 1\% were US election related. Of these 1\% of election related tweets, 0.74\% were Russian linked and had been detected by Twitter either as automation or spam.  In its report to Senate, Twitter only considers original Tweets; all retweets are excluded.

The report indicates that Twitter identified 131,000 tweets from the accounts identified in the IRA list. Of these 9\% were about the American Election (11,790). The total number of tweets annotated by Twitter as Election Related was 189 million which means that 0.006\% of Election related content was created by the IRA trolls. 

We analysed our longitudinal Brexit-related data set for evidence of activity from the 2,752 IRA linked troll accounts identified to Senate by Twitter. We confirmed that these accounts were creating Brexit related content. In total we found 3,485 tweets from the IRA linked accounts, representing 0.005\% of the total data we collected. 

The 0.005\% figure for Brexit as opposed to 0.006\% figure for the US Election indicates that there is a lower level of activity in our set of trolls discussing Brexit. Possible causal factors include: 
\begin{itemize}
\item that trolls, created to target the US election, are less active on the Brexit topic; 
\item trolls are more active at certain points, the lower level activity seen here could be a factor of the longitudinal nature of our data set. We gathered and analysed data from a period of over two years whilst Twitter presented an analysis of data from 1st September 2016 until 15th November 2016;
\item that there are other trolls that are more active in the Brexit debate but they are not on this list.
\end{itemize}

This 0.005\% figure includes retweets and drops to 0.002\% when these are excluded. In our troll dataset 57.59\% of data are retweets.

Our analysis of this longitudinal dataset allows us to present a study of the changes in activity and troll behaviour over time and in response to external events. In particular, we note a change in retweeting behaviour on the 23rd June 2016, the day of the UK-EU referendum vote. On this day we captured 1,059,888 tweets in total. Out of this total, 389 of these tweets were from troll accounts (0.037\%): nearly an eight-fold (7.4) increase in the relative number of tweets that came from trolls. The vast majority of tweets captured on this day were retweets, something the headline value of 0.74\% in the US Twitter report would fail to identify. In fact, only eleven tweets were original tweets. If we were to calculate troll activity excluding retweets, we find that on this day 0.001\% of original data are from trolls. Conducting the calculation in this way would indicate a decrease in troll activity rather than an increase. This highlights that, although the trolls were more active on the day of the Brexit referendum vote, there was a change in their behaviour and they produced more retweets and less original content. We must therefore consider this when we evaluate the 0.74\% value given by Twitter. 

 
\section{Brexit Data Collection}

We have been collecting Twitter data on the UK-EU referendum (Brexit) since August 2015. These data have enabled us to study discussions leading up the referendum and the consequential reaction to the decision of the UK to leave the European Union. Data were gathered through the Twitter API based on a selection of relevant hashtags chosen by a panel of academic experts. The set of hashtags grew periodically to reflect the evolving conversation. The dataset currently contains over seventy million tweets \cite{llewellyn2017brexit,llewellyn2016avoiding}. 

The terms of service of the Twitter Developer Agreement ask that all Tweets are `deleted within 24 hours after a request to do so by Twitter'. \footnote{\url{https://developer.twitter.com/en/developer-terms/agreement-and-policy}} We have an automated method in place that removes all tweets as requested. Therefore, it is possible tweets relevant to this study have been deleted.

This data is collected and archived to allow us to study the ongoing discussion and opinions regarding Brexit. The user accounts of the trolls identified by Twitter have now been deleted and are not available from Twitter directly. Archived copies of tweets are now the only way we can conduct academic research into troll activity in the Brexit discussion. We know that we are likely only to have a proportion of the content produced by the trolls. In this paper we can only analyse what we have found and remain aware that this data is probably a sub-sample of what was originally produced.

The selection of hashtags used to gather data ensured that we gathered both tweets that related to the Brexit vote directly and also to topics that were expected by experts to influence opinions on Brexit. When researching whether the troll accounts were active in the Brexit discussions we decided to split the data into tweets that were directly about Brexit and those that contained other Brexit related topics. 
We annotated tweets we had gathered from the troll accounts on the basis of whether they were directly about Brexit or not. The annotators were asked to be conservative and only to include tweets in the Brexit set if they were absolutely certain they were directly about Brexit. We found that 1,357 were directly about Brexit, 2,109 were not and 19 were difficult to decide. This gave us 38.94\% of the tweets that were directly about Brexit. Henceforth we will call these sets `Brexit Tweets' and `Related Tweets'. The 19 undecided tweets were excluded from the study. All of the tweets were annotated by a single coder. We double coded a sample (100) of the tweets to validate consistency and measure inter-annotator agreement, producing a kappa score of 0.80 indicative of very high agreement. These tweets contained multi-lingual content: English, German and Italian. Both annotators were fluent in all of these languages.
 
\section{Related Work} 
 Automatic generation of Twitter content is common-place. Bessi and Ferrara \shortcite{bessi2016social} found that one fifth of the Twitter conversation about the 2016 US Elections was not generated by humans.  `Bot' accounts are set up to automatically retweet and aggregate content from other sources or to create automatically generated text. `Influence bots' were described in the DARPA Twitter Bot Challenge as bots designed to influence discussion on social media sites \cite{subrahmanian2016darpa}.  Much of this content has no malicious intent, but some is designed to mislead or influence other users with misinformation and spam. 

Social media companies are not required to fact check information. Catchiness and repeatability can lead to widespread dissemination of content whether it is true or not \cite{ratkiewicz2011detecting}. Bots are often used as a method for repeating information and making it appear that the information is popular \cite{ferrara2016rise}. 

As bots have become more advanced, they are able to interact with other bots and humans in a conversational type way making them more believable and increasing their social networks \cite{ferrara2016rise}. Automatically extracting information from real users and from the wider internet allows the automatic generation of life-like user profile information creating complex and believable sock puppet personas \cite{ferrara2016rise}.

Automated accounts can be used to produce large amounts of content on single issues, where many accounts become active and tweet on the same topic at once forming a `bot legion' \cite{6280553}. Ratkiewicz et al \shortcite{ratkiewicz2011detecting} describe how nine fake users tweeted 929 times in 138 minutes in a 2009 Massachusetts election. This type of activity is intended to start a cascade of information-spreading with non-automated accounts reproducing the content. Messages are also more likely to be believed if they are seen from multiple sources \cite{ratkiewicz2011detecting} as it creates a wallpaper effect where information is seen so often it becomes background noise and is assumed to be true. 

As shown by Bessi and Ferrara \shortcite{bessi2016social} much of this automatically generated content can be automatically detected and extensive work has been conducted in this area, including the DARPA challenge. Systems have been created that are based on social network information, crowd sourcing and machine learning \cite{ferrara2016rise,davis2016botornot}. 

The 2,752 Twitter accounts detailed in the Senate list are not automated accounts but  'cyborgs' which are at least partially operated by humans \cite{chu2012detecting}. These are harder to detect than bots as they have the behaviour patterns of both bots and humans.

Astro-turfing  describes the use of bots, cyborgs or sock puppets to emulate the personas of individuals involved in grass root political movements \cite{ratkiewicz2011detecting}. This activity is intended to foster a sense of group identity amongst individuals that share certain traits. These traits help other users identify with them and make them more likely to be sympathetic towards their opinions. The more often human users are exposed to content the more likely they are to align with the perspectives being shared \cite{del2016spreading}.

Astro-turf cyborgs commonly combine political information with more general human content. Keller et al \shortcite{keller2017manipulate} found that cyborg trolls were used in astro-turfing by the South Korean secret service in the 2012 elections. They observed specific behaviour patterns: having many accounts tweet the same tweet at the same time to influence trending topics, having an agent cut and paste roughly the same content into many accounts, and a consistent time pattern for the activity in the accounts. They found that human troll accounts often act in similar and repetitive ways, as the individuals trolls are following central instructions. 

Howard and Kollanyi \shortcite{howard2016bots}, in a study of tweets collected between 5th and the 12th June 2016 in the UK-EU referendum, found that bots played a `small but strategic role in the referendum conversations' but that not all accounts were completely automated. Bastos and Mercea \shortcite{bastos2017brexit} found a network of 13,493 bots that tweeted on the UK-EU referendum but disappeared after the ballot, concluding that these accounts were involved in the amplification of human created content. 
    
\section{User Information}

The list provided by Twitter to the US Senate contained 2,752 accounts that were thought to contain troll activity. Within our full data set we found tweets from 419 of these accounts. We had more than one tweet from 66.83\% accounts. In the Brexit Tweets set we had tweets from 267 accounts. We found more than one tweet from 56.68\% of those accounts indicating that most accounts did engage with the Brexit topic multiple times.
\begin{table}[ht]\footnotesize
\centering
\caption{Most frequent Tweeters in all of the data (All), the data related to Brexit but not directly on Brexit (Related) and the data purely on Brexit (Brexit). Account usernames are anonymised.\\
}
\label{Freq}
\begin{tabular}{c|c|c|c|c|c}
\hline
\multicolumn{2}{c|}{All} & \multicolumn{2}{|c|}{Related} & \multicolumn{2}{|c}{Brexit}         \\ 
Account      & Tweets     & Account & Tweets & Account & Tweets \\ \hline
a            & 537        & a       & 375    & a       & 162    \\ \hline
b            & 280        & b       & 232    & b       & 48     \\ \hline
c            & 153        & d       & 109    & c       & 45     \\ \hline
d            & 131        & c       & 108    & e       & 30     \\ \hline
e            & 114        & e       & 84     & f       & 30     \\ \hline
f            & 95         & g       & 75     & l       & 29     \\ \hline
g            & 94         & f       & 64     & d       & 22     \\ \hline
h            & 71         & h       & 61     & m       & 21     \\ \hline
j            & 71         & j       & 58     & n       & 21     \\ \hline
k            & 45         & k       & 43     & o       & 20     \\ \hline
\end{tabular}
\end{table}

As seen in Table \ref{Freq} there is a high overlap between the most frequent tweeters in each set. The accounts tweeted both about Brexit and related topics. We have  most tweets from account  \textit{a}, an account that tweets in English and mostly retweets rather than producing original content.

\begin{figure}[]
\caption{Creation dates of the troll accounts.}
\label{accountDates}
\begin{tikzpicture}
	\begin{axis}[ybar stacked,
        ymin=0,
        ymax=120,
        symbolic x coords={2011, 2012, 2013, 2014, 2015, 2016, 2017},
        width=8cm,
        xtick=data,
        x tick label style={rotate=45,anchor=east}]
	\addplot coordinates
		{(2011,1) (2012,2) (2013,102) (2014,83) (2015,23) (2016,44) (2017,12)};
	\end{axis}
\end{tikzpicture}
\end{figure}
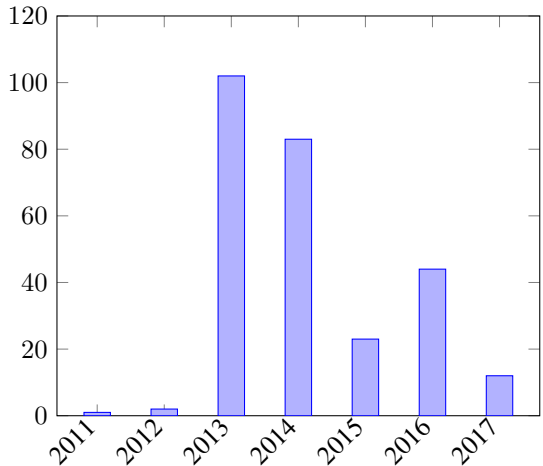

We examined the information from the user metadata as extracted from the troll tweets. This user metadata is either generated automatically or added by the account holder. The information can change over time and/or be altered by the user; for example the number of followers is an automated value that can vary but the user profile location field can be added by the account holder. Information that is added by the user can indicate the role of the account. If we assume that the accounts are all created by the Internet Research Agency, an account could be created to look like an American individual with a particular political opinion and this can be expressed through the information added in fields such as location and user description. 

Account creation date is an automatic field and can not be changed. As can be seen in Figure \ref{accountDates} most of the accounts were created in 2013 and 2014 after the 2012 US Election but well before both the 2016 US Election and the UK-EU Brexit Referendum. There are more accounts created in 2016, the year of the US Election (and the Brexit vote), than in either 2015 or 2017. Of the forty-four accounts created in 2016 thirty-eight were created after the UK Brexit referendum but before the US election, all but one of these within a very tight window of ten days between 4th and 13th of July. 

To further determine who the accounts were intended to represent we analysed the user description field in the tweet metadata. We counted the occurrences of terms by the year of account creation, and we removed very common english and german words. A user can change the text in this field at anytime but we did not find any evidence of changes in the data we collected. The full list of terms can be seen in Table \ref{user_terms}. Many of the accounts do not have any terms at all in the user description field. The counts in Table \ref{user_terms} are small but do indicate a pattern. In 2013-2015 the accounts contained description terms that indicate American, conservative, patriotic personas, suggesting that the accounts were designed to influence American events. But in 2016 many of the terms are German, mag (like), glaube (I believe), uern (likely a shortening of \"{a}u{\ss}ern which translates as express). Therefore, the accounts created in 2016 may not have been designed to tweet about the US election but something more European based instead.

When we looked at the user descriptions of the forty-four accounts created in 2016 we found twelve of the accounts created in July 2016 had German language descriptions (of the rest one was in English, one was mixed German and English, two only contained hashtags, and the rest were empty). For comparison, in 2015, sixteen were in English, three in German and four were empty.

This German language use in accounts created after the 2016 Brexit vote suggests that the trolls were using the result of the vote to push an agenda related to Germany, perhaps anticipating the German elections in 2017.

\begin{table*}[]
\centering
\caption{Frequency of terms from the user level description field split by year\\}
\label{user_terms}
\begin{tabular}{|l|l|l|l|l|l|l|l|l|l|}
\hline
\multicolumn{2}{|c|}{\textbf{2013}}  & \multicolumn{2}{|c|}{\textbf{2014}}  & \multicolumn{2}{|c|}{\textbf{2015}} & \multicolumn{2}{|c|}{\textbf{2016}} & \multicolumn{2}{|c|}{\textbf{2017}}  \\ \hline
conservative               & 16                    & conservative              & 9                     & love                      & 5                     & usa                       & 5                     & trump                     & 5                     \\ \hline
blacklivesmatter           & 15                    & tcot                      & 6                     & proud                     & 4                     & ttip                      & 3                     & 2a                        & 3                     \\ \hline
love                       & 11                    & wakeupamerica             & 5                     & tcot                      & 4                     & fr                        & 3                     & follow                    & 2                     \\ \hline
tcot                       & 11                    & patriot                   & 4                     & family                    & 4                     & mag                       & 3                     & mom                       & 2                     \\ \hline
dont                       & 8                     & supporter                 & 4                     & country                   & 4                     & glaube                    & 2                     & god                       & 2                     \\ \hline
pjnet                      & 7                     & life                      & 4                     & christian                 & 3                     & uern                      & 2                     & starke                    & 1                     \\ \hline
wakeupamerica              & 7                     & pjnet                     & 3                     & conservative              & 3                     & studiere                  & 2                     & corodinator               & 1                     \\ \hline
life                       & 7                     & dont                      & 3                     & patriot                   & 2                     & freizeit                  & 2                     & broker                    & 1                     \\ \hline
2a                         & 6                     & 2a                        & 3                     & youre                     & 2                     & spiele                    & 2                     & moment                    & 1                     \\ \hline
\end{tabular}
\end{table*}

We confirmed the change in location by checking by hand the location fields of tweets in the Brexit set. We did not see any evidence in our data set that these fields had been changed over time but this is possible. We found 154 were in some way based in the USA, sixteen in Europe and three in Russia (94 had no location information). The sixteen European accounts were made up of seven German accounts, five Italian accounts, three from the UK and one from Belgium. To give a better idea of the types of locations described we give the term counts from the location field in Table \ref{locTerm}. We can see that the European based accounts were created in 2016 and later. This location information suggests that most of the accounts on the list submitted by Twitter to US Senate were indeed designed to look like they are from the USA. The agenda they were designed to follow was also related to the USA, but those created in and after 2016 had a different agenda.

Some of the accounts have many followers and therefore a high potential to reach other Twitter users. We used the maximum number of followers when we had multiple tweets from an account. Of the 267 that have tweeted about Brexit: 122 accounts have more than 1,000 followers, sixteen accounts have over 10,000 and one account has over 100,000. The median number of followers is 875. As we cannot tell from this dataset how many of the trolls follow each other, this high median number should be treated with caution.

\begin{table*}[]
\centering
\caption{Frequency of terms from the user location field split by year\\}
\label{locTerm}
\begin{tabular}{|l|l|l|l|l|l|ll|l|l|}
\hline
\multicolumn{2}{|c|}{\textbf{2013}} & \multicolumn{2}{c|}{\textbf{2014}} & \multicolumn{2}{c|}{\textbf{2015}} & \multicolumn{2}{c|}{\textbf{2016}}   & \multicolumn{2}{c|}{\textbf{2017}} \\ \hline
usa                  & 51           & usa                  & 21          & usa                  & 6           & \multicolumn{1}{l|}{deutschland} & 3 & estados              & 4           \\ \hline
states               & 12           & atlanta              & 10          & texas                & 3           & \multicolumn{1}{l|}{berlin}      & 2 & unidos               & 4           \\ \hline
united               & 12           & us                   & 5           & germany              & 2           & \multicolumn{1}{l|}{hessen}      & 1 & italia               & 2           \\ \hline
chicago              & 4            & states               & 3           & brussel              & 1           &                                  &   & main                 & 1           \\ \cline{1-6} \cline{9-10} 
us                   & 4            & united               & 3           & stlouis              & 1           &                                  &   & lombardia            & 1           \\ \cline{1-6} \cline{9-10} 
il                   & 4            & la                   & 2           & tennessee            & 1           &                                  &   & frankfurt            & 1           \\ \cline{1-6} \cline{9-10} 
ny                   & 3            & new                  & 2           & states               & 1           &                                  &   & italy                & 1           \\ \cline{1-6} \cline{9-10} 
baltimore            & 2            & york                 & 2           & richmond             & 1           &                                  &   & sicilia              & 1           \\ \cline{1-6} \cline{9-10} 
ga                   & 2            & pittsburgh           & 1           & united               & 1           &                                  &   & itala                & 1           \\ \cline{1-6} \cline{9-10} 
atlanta              & 2            & ga                   & 1           & wisconsin            & 1           &                                  &   & milano               & 1           \\ \cline{1-6} \cline{9-10} 
\end{tabular}
\end{table*}

\section{Tweet Information}

The hashtags that are used by the troll accounts indicate the different topics discussed in both the Brexit Tweet and the Related Tweet sets. Table \ref{hashtagFreq} shows the top hashtags in each set. In the Brexit Tweet set the hashtags used are related to Brexit, Britain, and the EU. In the top ten we also find hashtags relating to Chancellor Merkel, \#merkel, and \#merkelmussbleiben (which translates as \#merkelmuststay). The trolls are directly discussing Brexit but also using wider hashtags for example referring to the role of the German Chancellor Merkel, underlining the wider European context of the Brexit debate. 

In the Related Tweet set we can see that the trolls use hashtags about the EU, \#eu; about refugees, \#refugeeswelcome, \#flchtlinge (which translates as \#refugee) and \#refugee. We see mentions of the German Chancellor, \#merkel and the President of Turkey \#erdogan, we also see reference to Germany, \#deutschland and Turkey \#trke. We also see that tweets that were classified as not directly about Brexit are still being tagged with the \#brexit hashtag. 

The way that the hashtags are being used in the wider Brexit related set suggests that the trolls have an agenda that related to Germany and Turkey and were using the Brexit topic to push this agenda and the issue of migration. The last elections in Germany were held on the 24th September 2017, and in Turkey there was a constitutional referendum held on the 16th April 2017.

\begin{table}[h]\footnotesize
\centering
\caption{Hashtags frequency across the data specifically on Brexit (Brexit) and those on Brexit related tweets (Related) \\}
\label{hashtagFreq}
\begin{tabular}{|l|l|l|l|}
\hline
\multicolumn{2}{|c|}{\textbf{Brexit}} & \multicolumn{2}{c|}{\textbf{Related}} \\ \hline
\#brexit             & 825  & \#eu                & 1206   \\ \hline
\#britaininout       & 378  & \#merkel            & 286    \\ \hline
\#euref              & 364  & \#refugeeswelcome   & 281    \\ \hline
\#brexitornot        & 211  & \#flchtlinge        & 199    \\ \hline
\#goodbyeuk          & 188  & \#erdogan           & 158    \\ \hline
\#brexitinout        & 186  & \#europe            & 153    \\ \hline
\#remainineu         & 168  & \#deutschland       & 146    \\ \hline
\#eu                 & 158  & \#trkei             & 128    \\ \hline
\#merkelmussbleiben  & 125  & \#brexit            & 113    \\ \hline
\#merkel             & 111  & \#refugee           & 80     \\ \hline
\end{tabular}
\end{table}

\subsection{Tweets, Retweets and Verified Users} 

We wanted to look at how the trolls behaved when directly discussing Brexit. We found that the Brexit Tweet subset contains a higher percentage of retweets and those retweets are less likely to be retweets of verified users and more likely to originally come from other trolls than the wider set. 

The total number of tweets that we gathered was 3,485; of these, 2,007 are retweets (57.59\%). Within the Brexit Tweet subset of 1,357 tweets 989 were retweets (72.88\%). In the full set there are 775 tweets that were originally created by verified users (38.61\%) and then retweeted by trolls. In the Brexit Tweet subset 321 tweets are originally from verified users (subsequently forming 32\% of retweets). 

Out of the 104 accounts retweeted in the Brexit Tweets set 57 (54.81\%) were only ever retweeted once. The most frequently retweeted verified users are presented in Table \ref{verified}. We can see that the top retweets come from news outlets, the EU Council and the EU Comission. 
\begin{table*}[t]\footnotesize
\centering
\caption{Verified Users Retweeted\\
}
\label{verified}
\begin{tabular}{|l|p{4cm}|p{1.5cm}|p{8cm}|}
\hline
\textbf{Account Retweeted} & \textbf{Screen Name}                        & \textbf{No. Retweets} & \textbf{Description} \\\hline
eucounciltvnews & EU Council TV NewsVerified account & 48 & Video coverage of Council news. Download our video packages in broadcast quality or embed them on your web page.                                           \\\hline
business        & Bloomberg                          & 40 & The first word in business news.                                                                                                                           \\\hline
eucopresident   & Donald Tusk                        & 13 & Twitter channel of Donald Tusk, President of the European Council. Managed by the media team.                                                              \\\hline
bloombergtv     & Bloomberg TV                       & 10 & Breaking news. Exclusive interviews. Market-moving scoops. Watch Bloomberg \#Daybreak LIVE on @Twitter, every weekday from 7:00 - 9:00 AM ET.              \\\hline
eu\_commission  & European Commission                & 10 & News and information from the European Commission. Tweets by the Social Media Team. Engaging on \#TeamJunckerEUpriorities: http://ec.europa.eu/priorities/ \\\hline
federicamog     & Federica Mogherini                 & 10 & High Representative of the EU for Foreign Affairs and Security Policy. Vice President of the EU Commission                                                 \\\hline
rtuknews        & RT UK                              & 8  & RT UK broadcasts from its studio in London. Watch LIVE UK news, documentaries and talk shows. Get even more on our website.                                \\\hline
zvezdanews      & TV channel "Zvezda"                & 7  & Nobody before us!                                                                                                                                          \\\hline
sputnikint      & Sputnik                            & 6  & Sputnik is a global wire, radio and digital news service. We exist to tell the stories that are not being told.                                            \\\hline
tagesschau      & tagesschau                         & 6  & The news of the ARD \\\hline                                                                                                                                      
\end{tabular}
\end{table*}

We also find that trolls retweet trolls. In the total set there are 443 retweets (22.07\%) from 34 different trolls (17 of these trolls were only in the set as retweets). In the Brexit Tweets set 345 are retweets (34.88\%) from 25 other trolls (10 only ever as retweets). In the Brexit Tweet subset one specific troll has been retweeted 186 times by the other trolls, which means that 5.33\% of all data gathered originated from a single account.
 
When we look at the Brexit Tweet set and split the data into original content and retweets from the troll accounts we can see a difference in the language used. We removed all of the hashtags, as these were the most frequent terms and masked other term usage. Very common english and a german words were also removed. If we look at Table \ref{termTweets} we can see that original content tends to contain german words whereas retweeted content tends to contain english terms.

\begin{table}[]\footnotesize
\centering
\caption{Terms from Brexit set tweets (hashtags removed), data split by whether the tweet is a retweet or not\\}
\label{termTweets}
\begin{tabular}{|l|l|l|l|l|l|}
\hline
 \multicolumn{2}{|c|}{\textbf{Original Content}} & \multicolumn{2}{c|}{\textbf{Retweets}} \\ \hline
nachdem                   & 32                 & eu                    & 62             \\ \hline
eu                        & 30                 & european              & 54             \\ \hline
fr                        & 25                 & uk                    & 41             \\ \hline
may                       & 21                 & doorstep              & 35             \\ \hline
gibt                      & 21                 & council               & 35             \\ \hline
merkel                    & 21                 & la                    & 28             \\ \hline
negativen                 & 20                 & deal                  & 26             \\ \hline
frau                      & 20                 & vote                  & 24             \\ \hline
folgen                    & 20                 & britain               & 24             \\ \hline
neue                      & 19                 & special               & 22             \\ \hline
\end{tabular}
\end{table}


\subsection{Tweet Dates}

\begin{figure}[]
\caption{Number of tweets per day that are directly relevant to brexit\\}
\label{freqDatesDay}
    \begin{tikzpicture}
        \begin{axis}[
            date coordinates in=x,
            xticklabel={\month.\year},
            yticklabel={}
            ]
            \addplot [mark=none, blue] table[x=datum,y=kommt] {zeiten.dat};
         
        \end{axis}
    \end{tikzpicture}
\end{figure}
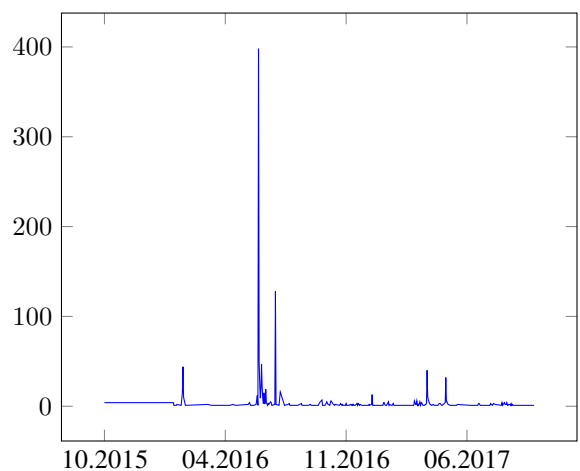

Overall there were very few tweets per day from the troll accounts. The number of tweets per day follows a similar pattern in both the Brexit Tweet and the Related Tweet sets. We can see in Figure \ref{freqDatesDay} that there are several spikes of activity for the Brexit Tweets dataset. In Table \ref{freqSpikes} we show the dates, volumes, whether the tweet was original content or a retweet and what was happening in the news that may have triggered the content. We found that the spikes in content production are related to particular events such as the referendum vote itself or the UK Prime Minister May meeting German Chancellor Merkel. We can see that there is a difference in troll behaviour on two dates that exhibit higher tweet volumes. On the 23rd June 2016 there is a large increase in volume of tweets produced but these are almost entirely retweets (97.73\%). In contrast, on the 21st July 2016 there was a considerably larger proportion of original content produced: only 26.56\% are retweets. There is also a conspicuous spike on the 19th February 2016. On a closer inspection of the data from this day, we found that all the tweets come from a single troll account.

\begin{table}[]\footnotesize
\centering
\caption{Further information about the days that had the highest frequency of tweets }
\label{freqSpikes}
\begin{tabular}{|l|l|l|p{3.5cm}|}
\hline
Date&Tweets&\%  RT & What happened on that day?\\\hline
19/02/2016                 & 44                                                                           & 100.00                                                                        & Cameron at EU summit                                                                           \\ \hline
23/06/2016                & 398                                                                          & 97.74                                                                      & UK-EU referendum                                                                               \\ \hline
24/06/2016                 & 51                                                                           & 49.02                                                                      & Day after UK-EU referendum                                                                     \\ \hline
28/06/2016                 & 47                                                                           & 70.21                                                                      &Cameron to meet EU leaders\\ \hline
21/07/2016                 & 128                                                                          & 26.56                                                                      & May meets Merkle                                                                               \\ \hline
29/03/2017                  & 40                                                                           & 45.00                                                                         & UK triggers Brexit                                                                             \\ \hline
29/04/2017                  & 32                                                                           & 100.00                                                                        & EU Discussion of brexit without UK                                                             \\ \hline
\end{tabular}
\end{table}

\section{The Day of the Brexit Referendum}

The largest number of troll tweets was collected on the day of the referendum, the 23rd of June 2016. We collected 400 in total of which 398 were directly about Brexit. On this day out of the 398 Brexit Tweets only nine tweets (and eleven tweets out of the 400 total) consisted of original content, the rest being retweets.  97.73\% of tweets on the day of the referendum were retweets. The trolls were therefore focused entirely on Brexit but they were retweeting rather than producing original content.

None of the tweets on the day of the UK's EU referendum were retweets of verified users. This is a radical change of behaviour. Out of the 387 retweets 279 were retweets of other trolls from the list isssued to Senate by Twitter (72.10\%). These tweets originate from only 11 troll accounts, and 186 were retweets originating from a single troll account. These 186 tweets exhibit a very similar homogeneity in style, content and format (text in italics altered from original):\\

\fbox{\begin{minipage}{20em}
\textit{@USER} \#brexitornot \#britaininout \#brexitinout \#euref https://t.co/\textit{VARIOUS}
\end{minipage}}\\\\

\begin{figure}[]
\caption{Number of tweets on the 23rd June by hour\\}
\label{freqDatesHour}
    \begin{tikzpicture}
        \begin{axis}[
            date coordinates in=x,
            xticklabel=\hour:\minute,
            xtick={
    {2016-06-23 06:00:00},
   {2016-06-23 10:00:00},
   {2016-06-23 14:00:00},
   {2016-06-23 18:00:00}
},
            yticklabel={}
            ]
            \addplot [mark=none, blue] coordinates{
(2016-06-23 06:00, 0)
(2016-06-23 07:00, 2)
(2016-06-23 08:00, 0)
(2016-06-23 09:00, 1)
(2016-06-23 10:00, 0)
(2016-06-23 11:00, 0)
(2016-06-23 12:00, 19)
(2016-06-23 13:00, 0)
(2016-06-23 14:00, 96)
(2016-06-23 15:00, 273)
(2016-06-23 16:00, 6)
(2016-06-23 17:00, 0)
(2016-06-23 18:00, 0)
(2016-06-23 19:00, 0)
(2016-06-23 20:00, 0)
 };        
        \end{axis}
    \end{tikzpicture}
\end{figure}
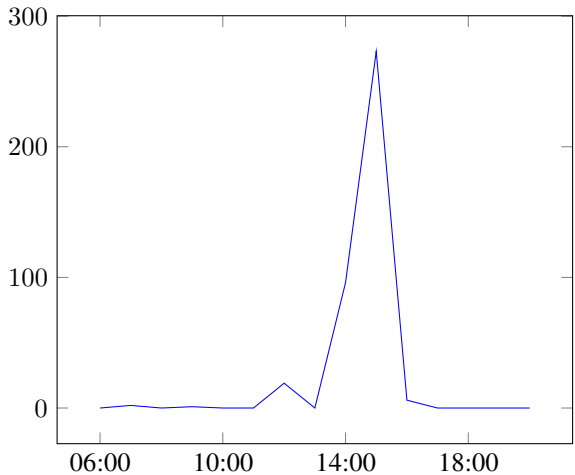

\begin{table}[]\footnotesize
\centering
\caption{The most frequent hashtags from the brexit dataset on the 23rd June 2016, the day of the referendum and on all other days\\ }
\label{brexitDayHash}
\begin{tabular}{|l|l|l|l|}
\hline
\textbf{All other days} & \textbf{} & \textbf{23rd June 2016} & \textbf{} \\ \hline
\#brexit                & 787       & \#britaininout          & 378       \\ \hline
\#eu                    & 148       & \#euref                 & 354       \\ \hline
\#merkelmussbleiben     & 125       & \#brexitornot           & 211       \\ \hline
\#merkel                & 110       & \#goodbyeuk             & 187       \\ \hline
\#euco                  & 89        & \#brexitinout           & 186       \\ \hline
\#ukineu                & 64        & \#remainineu            & 168       \\ \hline
\#may                   & 60        & \#brexit                & 38        \\ \hline
\#uk                    & 33        & \#reasonstoleaveeu      & 14        \\ \hline
\#article50             & 31        & \#eu                    & 10        \\ \hline
\#girlstalkselfies      & 18        & \#uk                    & 8         \\ \hline
\end{tabular}
\end{table}

In Figure \ref{freqDatesHour} we see the frequency of tweets grouped by hour across the day of the referendum. This reveals that the vast majority of tweets were sent between 2pm and 4pm. There were no tweets after 4pm although the referendum polls did not close until 10pm. The nine tweets consisting of original content were tweeted early in the day, while only two original tweets were tweeted after 2pm (at 2pm and 2.45pm).

\subsection{Amplification Behaviour}
A social media amplifier is defined as a user that shares ideas and opinions \cite{tinati2012identifying}. In this context we will use the term amplifier to classify an account which, as far as we can see from the data we have collected, only ever retweets. 

Overall there was very different pattern of behaviour exhibited by the trolls on the 23rd June, with a greater propensity to retweet, particularly sharing more information from other trolls. Overall in our set we have a large number of trolls who are amplifiers. To judge if their behaviour changed on the referendum polling day we analysed troll accounts which sent tweets on both the 23rd June and other days. In the Brexit Tweets set we have tweets from 248 troll accounts, of which 38 of them were active on the 23rd June 2016. Of those accounts, 19 (50\%) also appear on other days. This could suggest that the other 19 troll accounts only tweeted about Brexit on the 23rd June or it could just mean we did not catch them in our dataset.

There were only nine tweets that were not retweets on the 23rd June. Those original content tweets all came from accounts that tweeted on other days as well. In this data set the accounts that only tweeted on the 23rd June 2016 were amplifiers. Thirteen of the accounts acted as amplifiers on the 23rd June and twelve in the wider time period. As we do not have all of the tweets produced by all trolls we cannot establish a definitive pattern but this suggests that, while some accounts may simply be amplifiers, content producers can also switch their behaviour to amplification if required. Users may be more likely to be one or another but these behaviour patterns can change. On the day of the referendum vote we found that all of the IRA troll accounts were more likely to be involved in amplification behaviour.  

\section{Sentiment and Stance}

We annotated the 1,357 Brexit Tweets for both stance and sentiment. The annotator was asked to rate the stance of the tweets as either pro-leaving the EU, pro-remaining in the EU or neutral/neither. For sentiment the tweets were annotated as containing positive, negative or neutral sentiment. The majority of tweets were both neutral in stance (78.78\%) and neutral in sentiment (64.41\%), although this may be a consequence of the lack of available image or video context. In general the tweets had a stronger pro-leave stance (14.96\%) than pro-remain (6.26\%). The split of sentiment was fairly equal with a positive sentiment (18.35\%) being very slightly higher than negative sentiment (17.24\%). 

\begin{figure}[]
\caption{Sentiment Of Tweets\\}
\label{sent}
\begin{tikzpicture}
    \begin{axis}[
    ybar,
            date coordinates in=x,
            xticklabel=\year-\month,
            xtick={2016-01-01, 2016-06-01, 2016-12-01, 2017-06-01, 2017-12-01},
            yticklabel={},
            bar width=2pt,
ybar=0pt
      ]
        \addplot table[x=interval,y=neg]{mydata.dat};
        \addplot table[x=interval,y=pos]{mydata.dat};
        \legend{Negative, Positive}
    \end{axis}
\end{tikzpicture}
\end{figure}
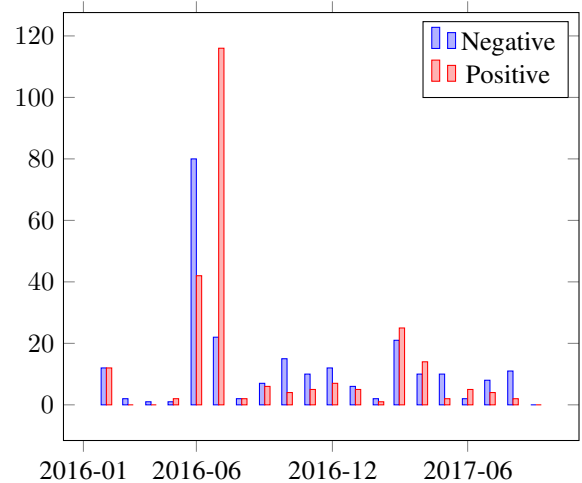

The pro-leave stance was consistently higher throughout the time period as shown in Figure \ref{stance}. The sentiment scores do change over time as can be seen in Figure \ref{sent}. In particular there was a spike of positive sentiment tweets on the 21st July 2016. As previously stated this was a spike in volume that occurred on the day that UK Prime Minister May met German Chancellor Merkel. A high percentage of these (73.44\%) were not retweets and the tweets were in German. The content driving this change in sentiment direction revolves around Chancellor Merkel, describing her as a strong person that will handle the Brexit issue well. The trolls discuss new possibilities and options after Brexit and that Frankfurt will be soon in a stronger position. A few trolls also talk about the EU accession of Turkey and that Merkel does not want any negotiations if Turkey re-introduces the death penalty.

\begin{figure}[]
\caption{Stance Of Tweets\\}
\label{stance}
\begin{tikzpicture}
    \begin{axis}[
    ybar,
            date coordinates in=x,
            xticklabel=\year-\month,
            xtick={2016-01-01, 2016-06-01, 2016-12-01, 2017-06-01, 2017-12-01},
            yticklabel={},
            bar width=2pt,
            ybar=0pt
      ]
        \addplot table[x=interval,y=leave]{mydatastance.dat};
        \addplot table[x=interval,y=remain]{mydatastance.dat};
        \legend{Pro-Leave, Pro-Remain}
    \end{axis}
\end{tikzpicture}
\end{figure}
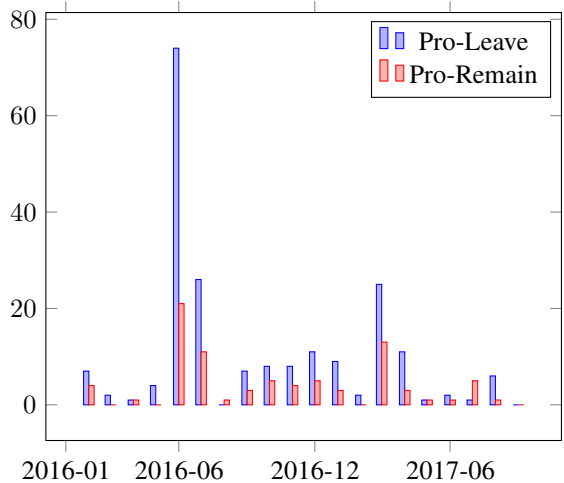

\section{User Accounts Released 19th Jan 2017}
On the 19th January 2017 Twitter released a statement saying that they were adding a further 1,062 accounts associated with the Internet Research Agency. We have yet to see these account usernames. 

\section{Discussion}
The Internet Research Agency accounts identified by Twitter were active in the Brexit debate. These tweets were slightly more likely to be pro-leave than pro-remain. 

These accounts were not designed to look like either pro-leave or pro-remain grassroots individuals. It is likely they were designed to be active in the American and perhaps latterly in the German Elections. This raises two questions: do accounts designed by the Internet Research Agency to look like grassroots Brexit groups exist? and why were the American and German sock-puppets tweeting about Brexit?

The answer to the first question is unknown. To the second question, it could be either be that these tweets are background noise designed to make the accounts look either more human or more politically aware, or that the topic of Brexit was used to promote another agenda such as instability and disruption. 

We also observe the high-level retweeting of verified users again possibly to create the illusion of a politically active human.

We observe `bot legion' behaviour on the 23rd June 2016. Many retweets that contain very similar content were tweeted over a short time frame. It is likely that these were produced automatically or exhibit the cut and paste behaviour seen in the South Korean Election. It looks like these American and German persona had been instructed to tweet on the Brexit topic en masse.

There are other spikes in data production that were not produced by mass retweet events. We see on the 21st July there was a high level of positive original content produced that wass related to Merkel and Brexit, certainly pointing towards the use of Brexit as a supporting issue. 

The longitudinal nature of the data collection technique that we use offers the opportunity to investigate behavioural changes and adaptation. If we had studied these data from the 23rd June 2016 in isolation then the cyborg troll accounts would have simply resembled bot accounts. In order to successfuly identify the cyborg accounts we need to systematically look for changes in behaviour over time.

\bibliographystyle{aaai}
\bibliography{trolls}
\end{document}